\documentclass{llncs}

\usepackage{makeidx}  
\usepackage{graphicx}
\DeclareGraphicsExtensions{.eps}
\usepackage[nolist]{acronym}
\usepackage[table]{xcolor}
\usepackage{listings}

\newcommand{\epsep}{\rceil}
\newcommand{\olsep}{\|}
\newcommand{\nolsep}{|}
\newcommand{\ecmspace}{\,}
\newcommand{\ecmshort}[5]{\mbox{$\left\{{#1}\ecmspace\olsep\ecmspace {#2}\ecmspace\nolsep\ecmspace {#3}\ecmspace\nolsep\ecmspace {#4}\ecmspace\nolsep\ecmspace {#5}\right\}$}}
\newcommand{\ecm}[6]{\mbox{$\left\{{#1}\ecmspace\olsep\ecmspace {#2}\ecmspace\nolsep\ecmspace {#3}\ecmspace\nolsep\ecmspace {#4}\ecmspace\nolsep\ecmspace {#5}\right\}\ecmspace{#6}$}}
\newcommand{\ecmp}[5]{\mbox{$\left\{{#1}\ecmspace\epsep\ecmspace {#2}\ecmspace\epsep\ecmspace {#3}\ecmspace\epsep\ecmspace {#4}\right\}\ecmspace{#5}$}}
\newcommand{\ecmpshort}[4]{\mbox{$\left\{{#1}\ecmspace\epsep\ecmspace {#2}\ecmspace\epsep\ecmspace {#3}\ecmspace\epsep\ecmspace {#4}\right\}$}}
\newcommand{\ecme}[5]{\mbox{${#1}\ecmspace\epsep\ecmspace {#2}\ecmspace\epsep\ecmspace {#3}\ecmspace\epsep\ecmspace {#4}\ecmspace{#5}$}}

\usepackage[font={small}]{caption, subfig}
\begin{document}


\begin{acronym}[CISC]
    \acro{AGU}{Address Generation Unit}
    \acro{AVX}{Advanced Vector Extensions}
    \acro{BPU}{Branch Prediction Unit}
    \acro{CISC}{Complex Instruction Set Computer}
    \acro{CL}{Cache Line}
    \acro{CoD}{Cluster on Die}
    \acro{DP}{Double Precision}
    \acro{ECM}{Execution-Cache-Memory}
    \acro{EDP}{Energy-Delay Product}
    \acro{FIVR}{Fully Integrated Voltage Regulators}
    \acro{FMA}{Fused Multiply-Add}
    \acro{Flops}{Floating-Point Operations}
    \acro{HA}{Home Agent}
    \acro{ILP}{Instruction Level Parallelism}
    \acro{IMCI}{Initial Many Core Instructions}
    \acro{ISA}{Instruction Set Architecture}
    \acro{LFB}{Line Fill Buffer}
    \acro{LLC}{Last-Level Cache}
    \acro{MC}{Memory Controller}
    \acro{NUMA}{Non-Uniform Memory Access}
    \acro{OoO}{Out-of-Order}
    \acro{PCIe}{Peripheral Component Interconnect Express}
    \acro{PRF}{Physical Register File}
    \acro{QPI}{QuickPath Interconnect}
    \acro{RFO}{Read for Ownership}
    \acro{RISC}{Reduced Instruction Set Computer}
    \acro{SIMD}{Single Instruction Multiple Data}
    \acro{SP}{Single Precision}
    \acro{SSE}{Streaming SIMD Extensions}
    \acro{UFS}{Uncore Frequency Scaling}
\end{acronym}

\mainmatter

\title{Analysis of Intel's Haswell Microarchitecture Using The ECM Model and Microbenchmarks}

\author{J. Hofmann\inst{1}, D. Fey\inst{1}, J. Eitzinger\inst{2}, G. Hager\inst{2} \and G. Wellein\inst{2}}

\authorrunning{Johannes Hofmann et al.}
\tocauthor{Johannes Hofmann, Dietmar Fey, Jan Eitzinger, Georg Hager, Gerhard Wellein}
\institute{Computer Architecture, University Erlangen--Nuremberg\\
\and
Erlangen Regional Computing Center (RRZE), University Erlangen--Nuremberg\\
\email{johannes.hofmann@fau.de}}

\maketitle

\begin{abstract}
This paper presents an in-depth analysis of Intel's Haswell microarchitecture for
streaming loop kernels. Among the new features examined is the
dual-ring Uncore design, Cluster-on-Die mode, Uncore Frequency Scaling, core
improvements as new and improved execution units, as well as
improvements throughout the memory hierarchy. The
Execution-Cache-Memory diagnostic performance model is used together with
a generic set of microbenchmarks to quantify the efficiency of the
microarchitecture. The set of microbenchmarks is chosen such that it can serve
as a blueprint for other streaming loop kernels.
\end{abstract}

\section{Introduction and Related Work}
In accord with Intel's tick-tock model, where a tick corresponds to a shrinking
of the process technology of an existing microarchitecture and a tock
corresponds to a new microarchitecture, Haswell is a tock and thus represents a
new microarchitecture. This means major changes to the preceding Ivy Bridge
release have been made that justify a thorough analysis of the new
architecture. This paper demonstrates how the \ac{ECM} diagnostic performance
model \cite{Hofmann:2015:1,Treibig:2009,hager:cpe,sthw15} can be used as a tool
to evaluate and quantify the efficiency of a microarchitecture.
The ECM model is a resource-centric model that allows to quantify the runtime
of a given loop kernel on a specific architecture. It requires detailed
architectural specifications and an instruction throughput prediction as input.
It assumes Perfect instruction level parallelism for instruction execution as
well as bandwidth-bound data transfers. As a consequence the model yields a
practical upper limit for single core performance. The only empirically
determined input for the model is that of sustained memory bandwidth, which can
be different for each benchmark.  The model quantifies different runtime
contributions from instruction execution and data transfers within the complete
memory hierarchy as well as potential overlap between contributions. Runtime
contributions are divided into two different categories: $T_\mathrm{nOL}$, i.e.
cycles in which the core executes instructions that forbid simultaneous
transfer of data between the L1 and L2 caches; and $T_\mathrm{OL}$, i.e. cycles
that do not contain non-overlapping instructions, thus allowing for
simultaneous instruction execution and data transfers between L1 and L2 caches.
Note that one improvement to the original \ac{ECM} model is that apart from
load instructions, store instructions are now also considered non-overlapping.
Instruction times as well as data transfer times, e.g. $T_\mathrm{L1L2}$
for the time required to transfer data between L1 and L2 caches, can be
summarized in shorthand notation:
\ecmshort{T_\mathrm{OL}}{T_\mathrm{nOL}}{T_\mathrm{L1L2}}{T_\mathrm{L2L3}}{T_\mathrm{L3Mem}}.
The in-core execution time $T_\mathrm{core}$ is the maximum of either
overlapping or non-overlapping instructions. Predictions for cache/memory
levels is given by $\max(T_\mathrm{OL}, T_\mathrm{nOL}+T_\mathrm{data})$ with
$T_\mathrm{data}$ the sum of the individual contributions up to the
cache/memory level under consideration, e.g. for the L3 cache
$T_\mathrm{data}=T_\mathrm{L1L2}+T_\mathrm{L2L3}$. A similar shorthand notation
exists for the model's prediction:
\ecmpshort{T_\mathrm{core}}{T_\mathrm{L2}}{T_\mathrm{L3}}{T_\mathrm{Mem}}{}.
For details on the ECM model refer to the previously provided references.

Related work covers in-detail analysis of architectural features using
microbenchmarks, e.g., \cite{Molka:2015,Molka:2014,Schone:2012}.  We are not
aware of any work though using an analytic model to quantify the efficiency of
a microarchitecture.

Section \ref{sec:intro_haswell}  presents major improvements in Intel Haswell.  In
Section \ref{sec:microbenchmarks} we introduce a comprehensive set of
microbenchmarks that serves as a blueprint for streaming loop kernels. To
evaluate the hardware, obtained measurements are correlated with the
performance predictions in Section \ref{sec:results}.

\section{Haswell Microarchitecture}
\label{sec:intro_haswell}

\subsection{Core Design}

\begin{figure}[tb]
\begin{minipage}{.49\textwidth}
    \centering
    \includegraphics[width=\linewidth]{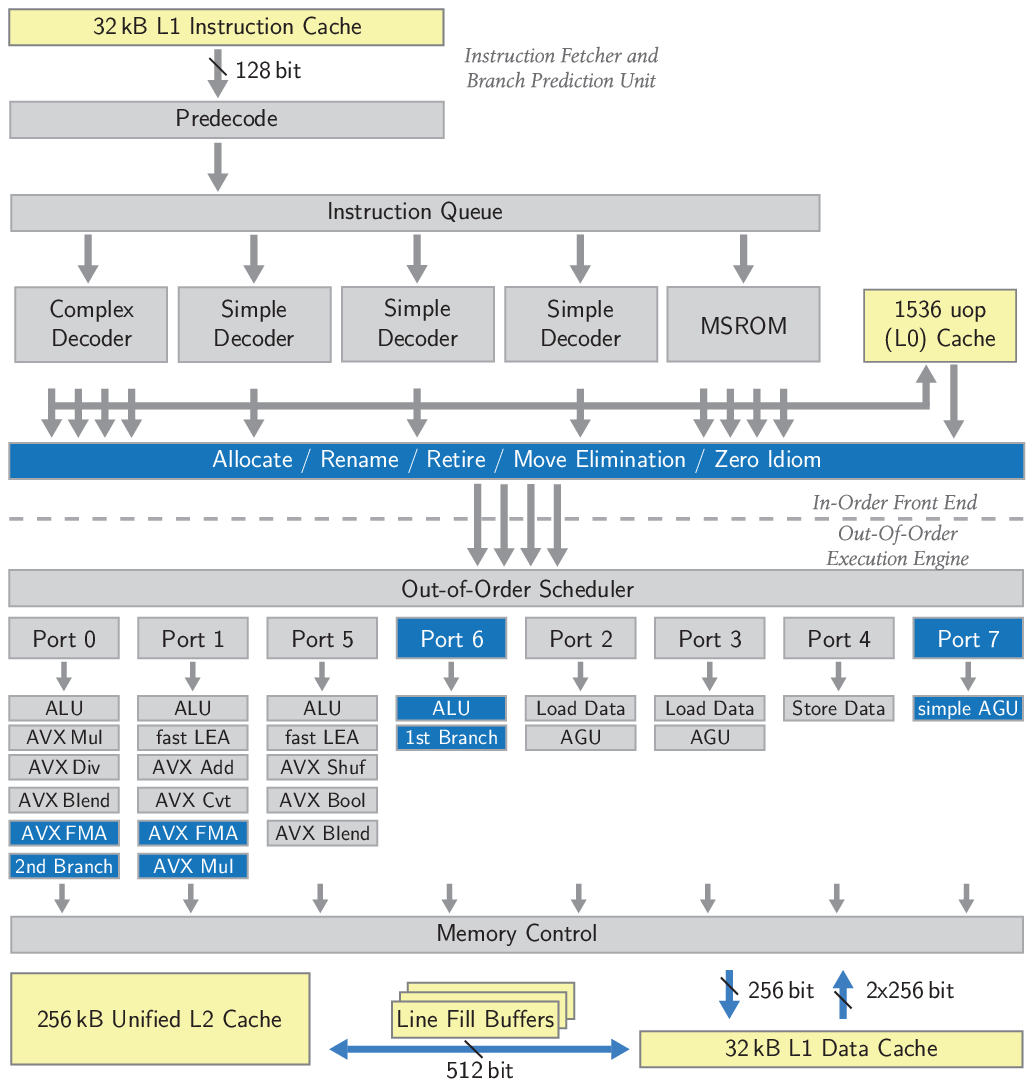}
    \caption{Core design for the Haswell Microarchitecture}
    \label{fig:hsw-pipeline}
\end{minipage}%
\hfill
\begin{minipage}{.49\textwidth}
    \centering
    \includegraphics[width=\linewidth]{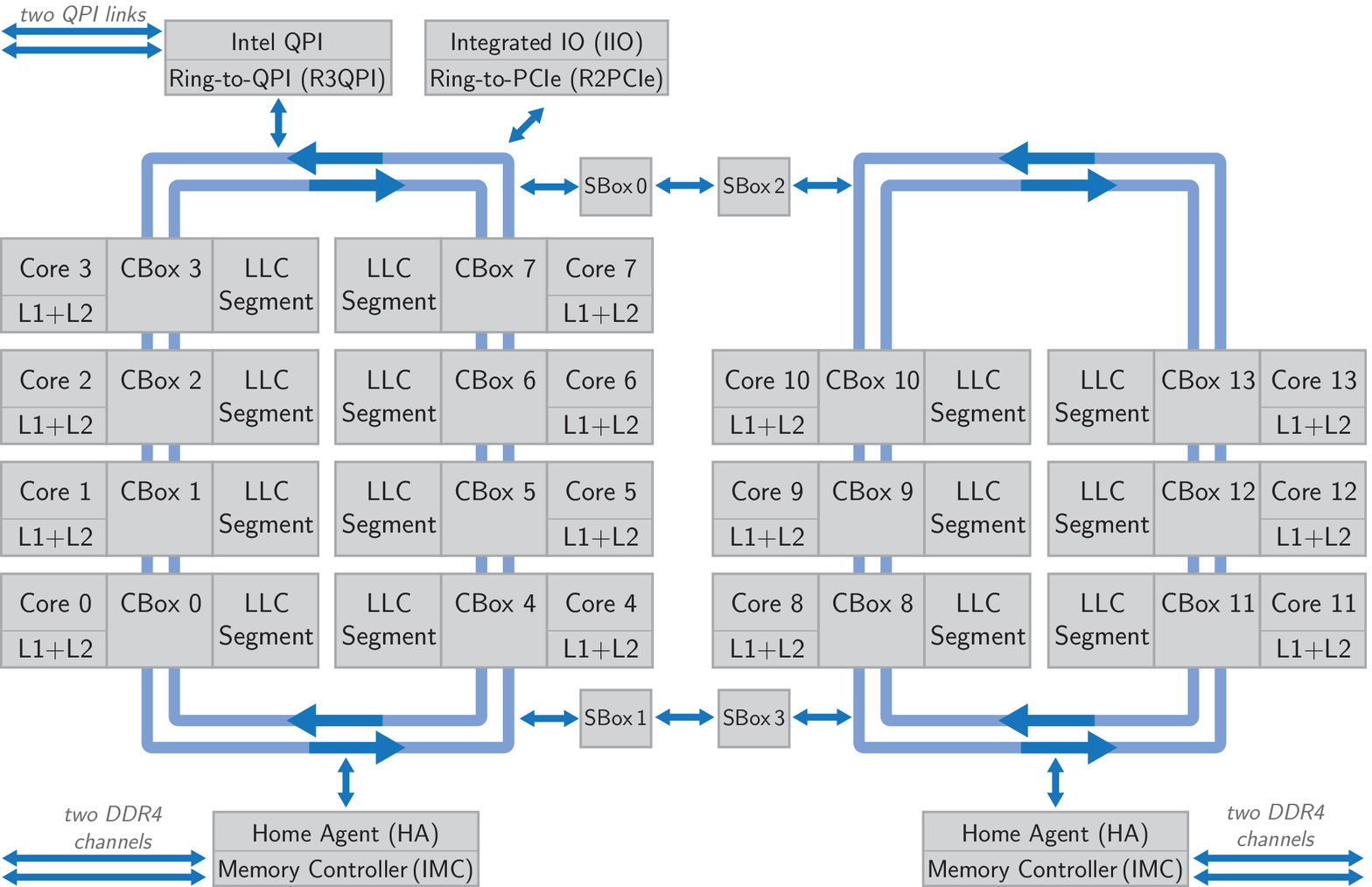}
    \caption{Chip layout for the Haswell Microarchitecture}
    \label{fig:hsw-package}
\end{minipage}
\end{figure}

Fig.~\ref{fig:hsw-pipeline} shows a simplified core design of the Haswell
microarchitecture with selected changes to previous
microarchitectures highlighted in blue. Due to lack of space we focus on
new features relevant for streaming loop kernels.

The width of all three data paths between the L1 cache and processor registers
has been doubled in size from 16\,B to 32\,B.  This means that two \ac{AVX} loads
and one store (32\,B in size) can now retire in a single clock cycle as opposed to
two clock cycles required on previous architectures. The data
path between the L1 and L2 caches has been widened from 32\,B to 64\,B.

While the core is still limited to retire four $\mu$ops per cycle, the number
of issue ports has been increased from six to eight. The newly introduced
port~6 contains the primary branch unit; a secondary unit has been added to
port~0. In previous designs only a single branch unit was available and located
on port~5.  By moving it to a dedicated port, port~5---which
is the only port that can perform AVX shuffle operations---is freed up. Adding
a secondary branch unit benefits branch-intensive codes. The other new port
is port~7, which houses a so-called simple \ac{AGU}. This unit was made
necessary by the increase in register-L1 bandwidth. Using AVX on Sandy Bridge
and Ivy Bridge, two \ac{AGU}s were sufficient, because each load or store
required two cycles to complete, not making it necessary to compute three new
addresses every cycle, but only every second cycle. With Haswell this has
changed, because potentially a maximum of three load/store operations can now
retire in a single cycle, making a third \ac{AGU} necessary. Unfortunately,
this simple \ac{AGU} can not perform the necessary addressing operations
required for streaming kernels on its own (see Section~\ref{sec:res:triads} for
more details).

Apart from adding additional ports, Intel also extended existing ones with new
functionality. Instructions introduced by the \ac{FMA} \ac{ISA} extension are
handled by two new, AVX-capable units on ports~0 and~1. Haswell is the
first architecture to feature the AVX2 ISA extension and introduces
a second \ac{AVX} multiplication unit on port~1 while there is still just one
low-latency add unit.

\subsection{Package Layout}
\label{sec:hsw:package}

Figure~\ref{fig:hsw-package} shows the layout of a 14-core Haswell processor
package. Apart from the processor cores, the package consists of what Intel
refers to as the Uncore. Attached to each core and its private L1 and L2
caches, there is a \ac{LLC} segment, that can hold 2.5\,MB of data. The
physical proximity of core and cache segment does however not imply that data
used by a core is stored exclusively or even preferably in its \ac{LLC} segment.
Data is placed in all LLC segments according to a proprietary hash function
that is supposed to provide uniform distribution of data and prevent hotspots for
a wide range of data access patterns. An added benefit of this design is that
single-threaded applications can make use of the full accumulated \ac{LLC} capacity.

The cores and \ac{LLC} segments are connected to a bidirectional ring
interconnect that can transfer one \ac{CL} (64\,B in size) every two cycles in
each direction.  In order to reduce latency, the cores are arranged to form two
rings, which are connected via two queues. To each ring belongs a
\ac{HA} which is responsible for cache snooping operations and reordering of
memory requests to optimize memory performance. Attached to each \ac{HA} is a
\ac{MC}, each featuring two 8\,byte-wide DDR4 memory channels.  Also accessible
via the ring interconnect are the on-die PCIe and QPI facilities.

Haswell introduces an on-die \ac{FIVR}. This \ac{FIVR}
draws significantly less power than on previous
microarchitectures, because it enabled for faster switching of
power-saving states. It also allows a more fine-grained control of CPU states:
instead of globally setting the CPU frequencies for all cores within a package,
Haswell can now set core frequencies and sleep states individually.

\subsection{Uncore Frequency Scaling}

\begin{figure}[tb]
\begin{minipage}{.49\textwidth}
    \centering
    \includegraphics[width=\linewidth]{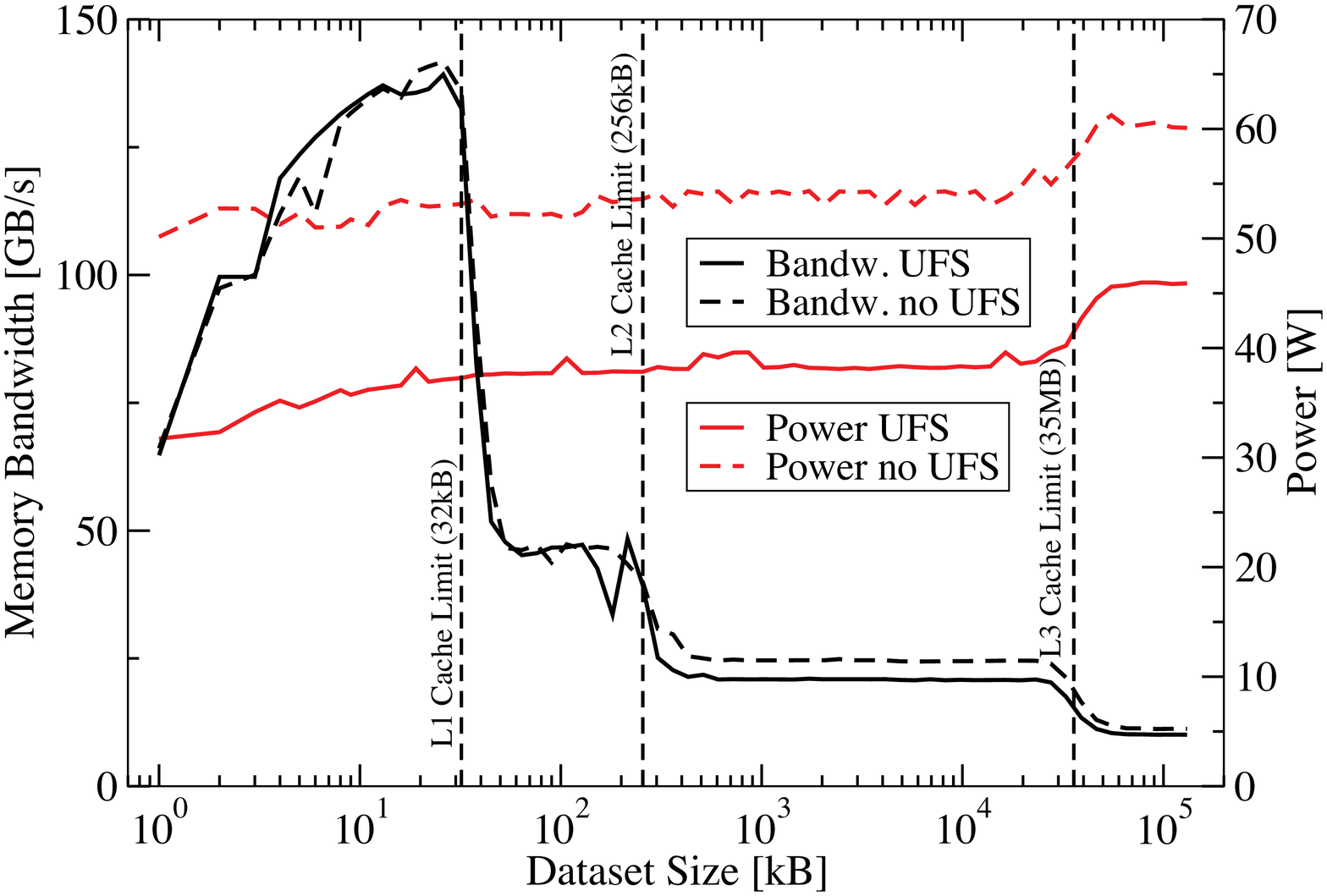}
    \caption{Impact of \ac{UFS} on Bandwidth and Power Usage.}
    \label{fig:hsw-ufs}
\end{minipage}%
\hfill
\begin{minipage}{.49\textwidth}
    \centering
    \includegraphics[width=\linewidth]{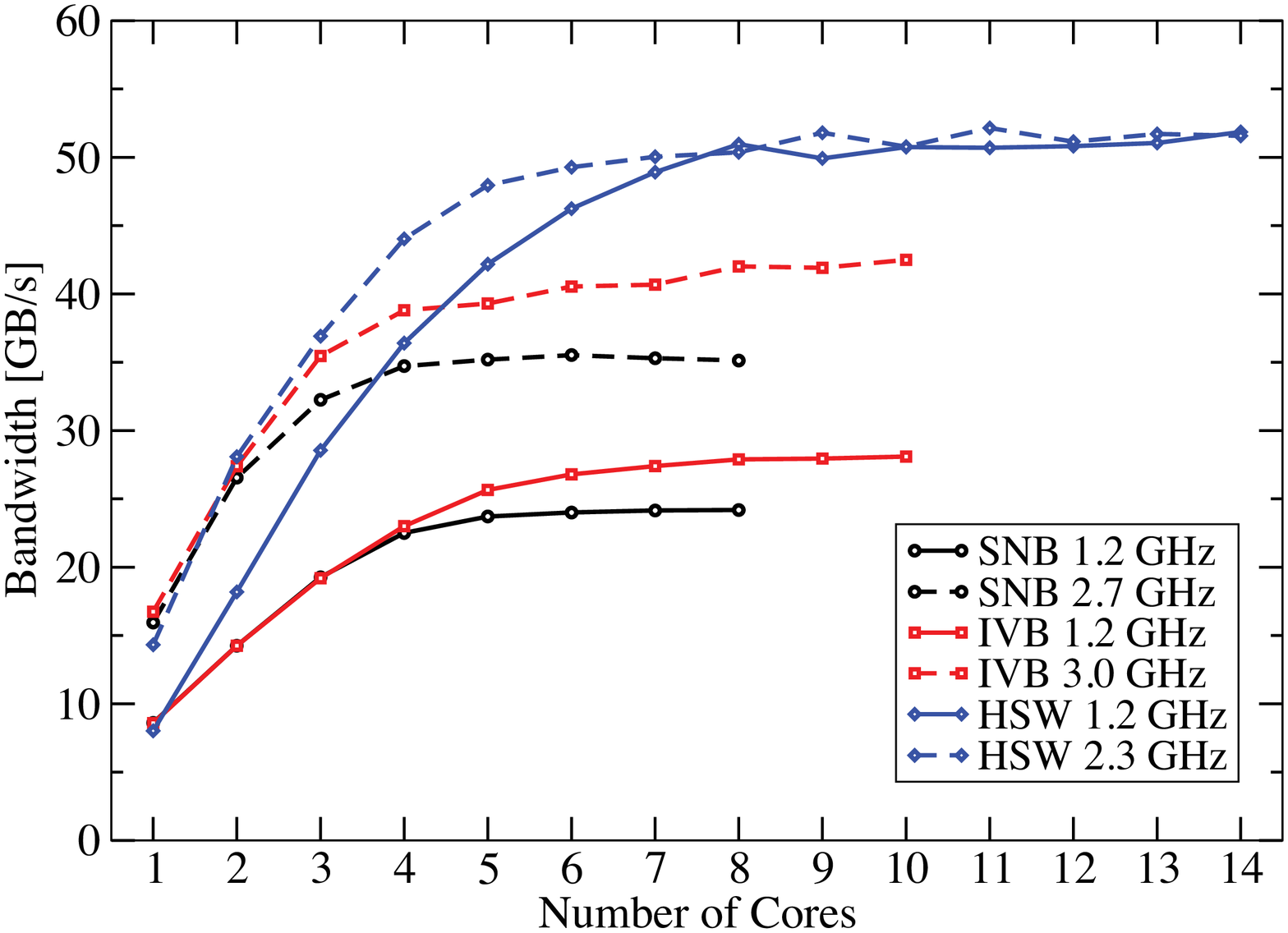}
    \caption{Stream Triad Bandwidth as Function of Frequency.}
    \label{fig:bw-freq-comparison-stream-triad}
\end{minipage}
\end{figure}

In the new Haswell microarchitecture, Intel reverted from Sandy and Ivy
Bridges' unified clock domain for core and the Uncore to the Nehalem design of
having two separate clock domains
\cite{Intel:xeon-e5-v1-overview,Intel:techjournal}.  Haswell introduced a
feature called \ac{UFS}, in which the Uncore frequency is dynamically scaled
based on the number of stall cycles in the CPU cores.  Despite reintroducing
higher latencies, the separate clock domain for the Uncore offers a significant
potential for power saving, especially for serial codes.
Fig.~\ref{fig:hsw-ufs} shows the measured sustained bandwidth (left $y$-axis)
for the Sch\"on\-au\-er vector triad (cf. Table~\ref{tab:benchmarks}) using a
single core along with the power consumption (right $y$-axis) for varying
dataset sizes. As expected the performance is not influenced by whether UFS is
active or not when data resides in a core's private caches; however, power
requirements are reduced by about 30\%!  Although we observe a difference in
performance as soon as the \ac{LLC} is involved, the performance impact is
limited. The bandwidth drops from 24 to 21\,GB/s (about 13\%) in the \ac{LLC},
but power usage is reduced from 55\,W to 40\,W (about 27\%).

\subsection{Memory}

Intel's previous microarchitectures show a strong correlation between
CPU frequency and sustained memory bandwidth.
Fig.~\ref{fig:bw-freq-comparison-stream-triad} shows the
measured chip bandwidth for the Stream Triad (cf.
Table~\ref{tab:benchmarks})---adjusted by a factor of 1.3 to account for the
write-allocate when storing---on the Sandy Bridge, Ivy Bridge, and Haswell
microarchitectures.  For each system, the bandwidth was measured using the
lowest possible frequency (1.2\,GHz) and the advertised nominal
clock speed. While we find differences around 33\% in the maximum achievable
sustained memory bandwidth depending on the CPU frequency for Sandy and Ivy
Bridge, on Haswell we can observe a frequency-independent sustained bandwidth
of 52.3\,GB/s. On Haswell the CPU frequency can be lowered, thereby decreasing
power consumption, while the memory bandwidth stays constant.
Further research regarding the Stream Triad with a working set size of 10\,GB has shown that
Haswell offers an improvement of 23\% respectively 12\% over the Sandy and Ivy
Bridge architectures when it comes to energy consumption and 55\% respectively
35\% in terms of Energy-Delay Product \cite{Hofmann:2015:1}.
\subsection{Cluster on Die}

In \ac{CoD} mode, cores get equally separated into two ccNUMA memory domains.
This means that instead of distributing requests between both memory
controllers each core is assigned a dedicated memory controller. To keep
latencies low, the strategy is to make a core access main memory through the
memory controller attached to its ring. However, with memory domains being
equal in size, the asymmetric core count on the two physical rings makes
exceptions necessary. In the design shown in Fig.~\ref{fig:hsw-package} the 14
cores are divided into two memory domains of 7 cores each.  Using
microbenchmarks and \texttt{likwid-perfctr} \cite{Treibig:2011:3} to access
performance counters in order to measure the number of memory accesses for each
individual memory channel, we find that cores 0--6 access main memory through
the memory channels associated with the memory controller on the left ring, and
cores 7--13 those associated with the memory controller on ring 1. Thus, only
core number 7 has to take a detour across rings to access data from main
memory. Note that with \ac{CoD} active the \ac{LLC} also is divided. As
each domain contains seven \ac{LLC} segments (2.5\,MB each), the total amount
of \ac{LLC} for each domain is only 17.5\,MB instead of 35\,MB.

\ac{CoD} mode is intended for NUMA-optimized codes and serves two purposes:
First latency is decreased  by reducing the number of endpoints in the memory
domain. Instead of 14 LLC segments, data will be distributed in only 7 segments
inside each memory domain, thereby decreasing the mean hop count.  Also, the
requirement to pass through the two buffers connecting the rings is eliminated
for all but one \ac{LLC} segment.  Also, bandwidth is increased by reducing the
probability of ring collisions by lowering participant count from 14 to~7.

\section{Microbenchmarks}
\label{sec:microbenchmarks}

A set of microbenchmarks chosen to provide a good coverage of relevant data
access patterns was used to evaluate the Haswell microarchitecture and is
summarized in Table~\ref{tab:benchmarks}.  For each benchmark, the table lists
the number of load and store streams---the former being divided into explicit
and \ac{RFO} streams. \ac{RFO} refers to implicit loads that occur whenever a
store miss in the current cache triggers a write-allocate. On Intel
architectures all cache levels use a write-allocate strategy on store misses.
The table also includes the predictions of the ECM model and the actually
measured runtimes in cycles along with a quantification of the model's error.
In the following, we discuss the ECM model for each of the kernels and show how
to arrive at the prediction shown in the table.

The sustained bandwidths used to derive the L3-memory cycles per \ac{CL} inputs
can be different for each benchmark, which is why for each individual kernel
the sustained bandwidth is determined using a benchmark with the exact data
access pattern that is modeled; note that for our measurements \ac{CoD} mode
was active and the measured bandwidth corresponds to that of a single memory
domain.

\begin{table*}[tb]
\renewcommand{\arraystretch}{1.2}      
\caption{Overview of microbenchmarks: Loop Body, Memory Streams, ECM prediction
and Measurement in c/CL, and Model Error.}
\label{tab:benchmarks}
\centering
\rowcolors{2}{gray!20}{white}
\resizebox{\linewidth}{!}{ 
\begin{tabular}{llccccc}
\hline
                    &                           & Load Streams      & Write     & ECM Prediction            & Measurement                      & Model Error \\
    Benchmark       & Description               & Explicit / RFO    & Streams   & L1/L2/L3/Mem              & L1/L2/L3/Mem                     & L1/L2/L3/Mem \\
\hline
    ddot            & \texttt{s+=A[i]*B[i]}     & 2 / 0             & 0         & \ecmp{2}{4}{8}{17.1}{}    & \ecme{2.1}{4.7}{9.6}{19.4}{}     & \ecme{5\%}{17\%}{20\%}{13\%}{} \\
    load            & \texttt{s+=A[i]}          & 1 / 0             & 0         & \ecmp{2}{2}{4}{8.5}{}     & \ecme{2}{2.3}{5}{10.5}{}         & \ecme{0\%}{15\%}{25\%}{23\%}{} \\
    store           & \texttt{A[i]=s}           & 0 / 1             & 1         & \ecmp{2}{4}{8}{20.5}{}    & \ecme{2}{6}{8.2}{17.7}{}         & \ecme{0\%}{33\%}{3\%}{16\%}{} \\
    update          & \texttt{A[i]=s*A[i]}      & 1 / 0             & 1         & \ecmp{2}{4}{8}{20.5}{}    & \ecme{2.1}{6.5}{8.3}{17.6}{}     & \ecme{5\%}{38\%}{4\%}{16\%}{} \\
    copy            & \texttt{A[i]=B[i]}        & 1 / 1             & 1         & \ecmp{2}{5}{11}{27.8}{}   & \ecme{2.1}{8}{13}{27}{}          & \ecme{5\%}{38\%}{15\%}{3\%}{} \\
    STREAM triad    & \texttt{A[i]=B[i]+s*C[i]} & 2 / 1             & 1         & \ecmp{3}{7}{15}{36.7}{}   & \ecme{3.1}{10}{17.5}{37}{}       & \ecme{3\%}{30\%}{14\%}{1\%}{} \\
    Sch\"onauer triad& \texttt{A[i]=B[i]+C[i]*D[i]} & 3 / 1         & 1         & \ecmp{4}{9}{19}{45.5}{}   & \ecme{4.1}{11.9}{21.9}{46.8}{}   & \ecme{3\%}{24\%}{13\%}{3\%}{} \\
\hline
\end{tabular}
} 
\end{table*}

\subsection{Dot Product and Load}
The dot product benchmark \textit{ddot} makes use of the new \ac{FMA}
instructions introduced in the FMA3 ISA extension implemented in the
Haswell microarchitecture.  $T_\mathrm{nOL}$ is two clock cycles, because the
core has to load two CLs (\texttt{A} and \texttt{B}) from L1 to registers using
four AVX loads (which can be processed in two clock cycles, because each
individual AVX load can be retired in a single clock cycle and there are two
load ports). Processing data from the CLs using two AVX \ac{FMA} instructions
only takes one clock cycle, because both issue ports~0~and~1 feature AVX FMA
units. A total of two CLs has to be transfered between the adjacent cache
levels. At 64\,B/c this means 2\,c to transfer the CLs from L2 to L1.
Transferring the CLs from L3 to L2 takes 4\,c at 32\,B/c. The empirically
determined sustained (memory domain) bandwidth is 32.4\,GB/s. At 2.3\,GHz, this
corresponds to a bandwidth of about
$64\,\mathrm{B/CL} \cdot 2.3\,\mathrm{GHz} / 32.4\,\mathrm{GB/s} \approx
4.5\,\mathrm{c/CL}$
or 9.1\,c for two CLs. The
ECM model input is thus \ecm{1}{2}{2}{4}{9.1}{\mathrm{c}} and the
corresponding prediction is \ecmp{2}{4}{8}{17.1}{\mathrm{c}}.

For the \textit{load} kernel the two AVX loads to get the CL containing
\texttt{A} from L1 can be retired in a single cycle, yielding
$T_\mathrm{nOL}$=1\,c. With only a single AVX add unit available on port~1,
processing the data takes $T_\mathrm{OL}$=2\,c.  Because only a single CL has
to be transferred between adjacent cache levels and the measured bandwidth
corresponds exactly to that of the \textit{ddot} kernel, the time required is
exactly half of that needed for the \textit{ddot} benchmark.  The ECM model
input for this benchmark is \ecm{2}{1}{1}{2}{4.5}{\mathrm{c}}, yielding a
prediction of \ecmp{2}{2}{4}{8.5}{\mathrm{c}}.

\subsection{Store, Update, and Copy}

For the \textit{store} kernel, two AVX stores are required per CL. With only a
single store unit available, $T_\mathrm{nOL} = 2$\,c; as there are no other
instructions such as arithmetic operations, $T_\mathrm{nOL}$ is zero. When
examining CL transfers along the cache hierarchy, we have to bear in mind that
a store-miss will trigger a write-allocate, resulting in two CL transfers for
each CL update: one to write-allocate the CL which data gets written to and one
to evict the modified CL once the cache becomes full.  This results in a
transfer time of 2\,c to move the data between the L1 and L2 cache and a
transfer time of 4\,c for L2 and L3. The sustained bandwidth of 23.6\,GB/s
(corresponding to approximately 6.2\,c/CL) for a kernel involving evictions is
significantly worse than that of the previous load-only kernels. The resulting
ECM input and prediction are \ecm{0}{2}{2}{4}{12.5}{\mathrm{c}} respectively
\ecmp{2}{4}{8}{20.5}{\mathrm{c}}.

For the \textit{update} kernel, two AVX stores and two AVX loads are required.
Limited by a single store port, $T_\mathrm{nOL} = 2$\,c.  The multiplications
take $T_\mathrm{OL} = 2$\,c.\footnote{Normally, with two AVX mul ports
available, $T_\mathrm{OL}$ should be 1\,c. However, the frontend can only
retire 4 $\mu$ops/c; this, along with the fact that stores count as 2 $\mu$ops,
means that if both multiplications were paired with the first store, there
would not be enough full AGUs to retire the second store and the remaining AVX
load instructions in the same cycle.} The number of CL transfers is
identical to that of the \textit{store} kernel, the only difference being that
the CL load is caused by explicit loads and not a write-allocate. With a memory
bandwidth almost identical to that of the \textit{store} kernel, the time to
transfer a CL between L3 and memory again is approximately 6.2\,c/CL, yielding
an ECM input of \ecm{2}{2}{2}{4}{12.5}{\mathrm{c}} and a prediction that
is identical to that of the \textit{store} kernel.

The \textit{copy} kernel has to perform two AVX loads and two AVX stores to
copy one CL. The single store port is the bottleneck, yielding
$T_\mathrm{nOL}=2$\,c; absent arithmetic instructions $T_\mathrm{nOL}$ is zero.
Three CLs have to be transferred between adjacent cache levels: load
\texttt{B}, write-allocate and evict \texttt{A}. This results in a requirement
of 3\,c for L1--L2 transfers and 6\,c for L2--L3 transfers.  With a sustained
memory bandwidth of 26.3\,GB/s the time to transfer one CL between main memory
and \ac{LLC} is approximately 5.6\,c/CL or 16.8\,c for three CLs.  This results
in the following input for the ECM model \ecm{0}{2}{3}{6}{16.8}{\mathrm{c}},
which in turn yields a prediction of \ecmp{2}{5}{11}{27.8}{\mathrm{c}}.

\subsection{Stream Triad and Sch\"onauer Triad}

For the \textit{STREAM Triad} \cite{McCalpin:1995}, the \ac{AGU}s prove to be
the bottleneck: it is impossible to retire two AVX loads and an AVX store that
use indexed addressing in the same cycle, because there are only two full
\ac{AGU}s available supporting this addressing mode.  The resulting
$T_\mathrm{nOL}$ thus is not 2 but 3\,c to issue four AVX loads (two each for
CLs containing \texttt{B} and \texttt{C}) and two AVX stores (two for CL
\texttt{A}). Both \ac{FMA}s can be retired in one cycle, because two \ac{AVX}
\ac{FMA} units are available, yielding $T_\mathrm{OL}$=1\,c. Traffic between
adjacent cache levels is 4 CLs: load CLs containing \texttt{B} and \texttt{C},
write-allocate and evict the CL containing \texttt{A}. The measured sustained
bandwidth of 27.1\,GB/s corresponds to approximately 5.4\,c/CL---or about
21.7\,c for all four CLs.  The input parameters for the ECM model are thus
\ecm{1}{3}{4}{8}{21.7}{\mathrm{c}} leading to the follow prediction:
\ecmp{3}{7}{15}{36.7}{\mathrm{c}}.

For the \textit{Sch\"onauer Triad} \cite{schoenauer:00}, again the AGUs are the
bottleneck. Six AVX loads (CLs \texttt{B}, \texttt{C}, and \texttt{D}) and two
AVX stores (CL \texttt{A}) have to be performed; these eight instructions have
to share two AGUs, resulting in $T_\mathrm{nOL}$=4\,c. The two AVX FMAs can be
performed in a single cycle, yielding $T_\mathrm{OL}$=1\,c.  Data transfers
between adjacent caches correspond to five CLs: \texttt{B}, \texttt{C}, and
\texttt{D} require loading while CL \texttt{A} needs to be write-allocated and
evicted. For the L1 cache, this results in a transfer time of 5\,c. The L2
cache transfer time is 10 cycles. The measured sustained memory bandwidth of
27.8\,GB/s corresponds to about 5.3\,c/CL or 26.5\,c for all five CLs.  The
resulting ECM input parameters are thus \ecm{1}{4}{5}{10}{26.5}{\mathrm{c}} and
the resulting prediction is \ecmp{4}{9}{19}{45.5}{\mathrm{c}}.


\section{Results}
\label{sec:results}

The results presented in this section were obtained using hand-written assembly
kernels that were benchmarked using the \texttt{likwid-bench} tool
\cite{Treibig:2011:3}.  No software prefetching was used in the code; the
results therefore show the ability of the hardware prefetchers to hide data
access latencies. The machine used for benchmarking was a standard two-socket
server using Xeon E5-2695\,v3 chips, featuring 14 cores each. Each core comes
with its own 32\,kB private L1 and 256\,kB private L2 caches; the shared
\ac{LLC} is 35\,MB in size. Each chip features four DDR4-2166 memory channels,
adding up to a theoretical memory bandwidth of 69.3\,GB/s per socket or
138.6\,GB/s for the full node.  For all benchmarks, the clock frequency was
fixed at the nominal frequency of 2.3\,GHz, \ac{CoD} was activated, and
\ac{UFS} was disabled.

\subsection{Dot Product and Load}

\begin{figure}[tb]
\begin{minipage}{.49\textwidth}
    \centering
    \includegraphics[width=\linewidth]{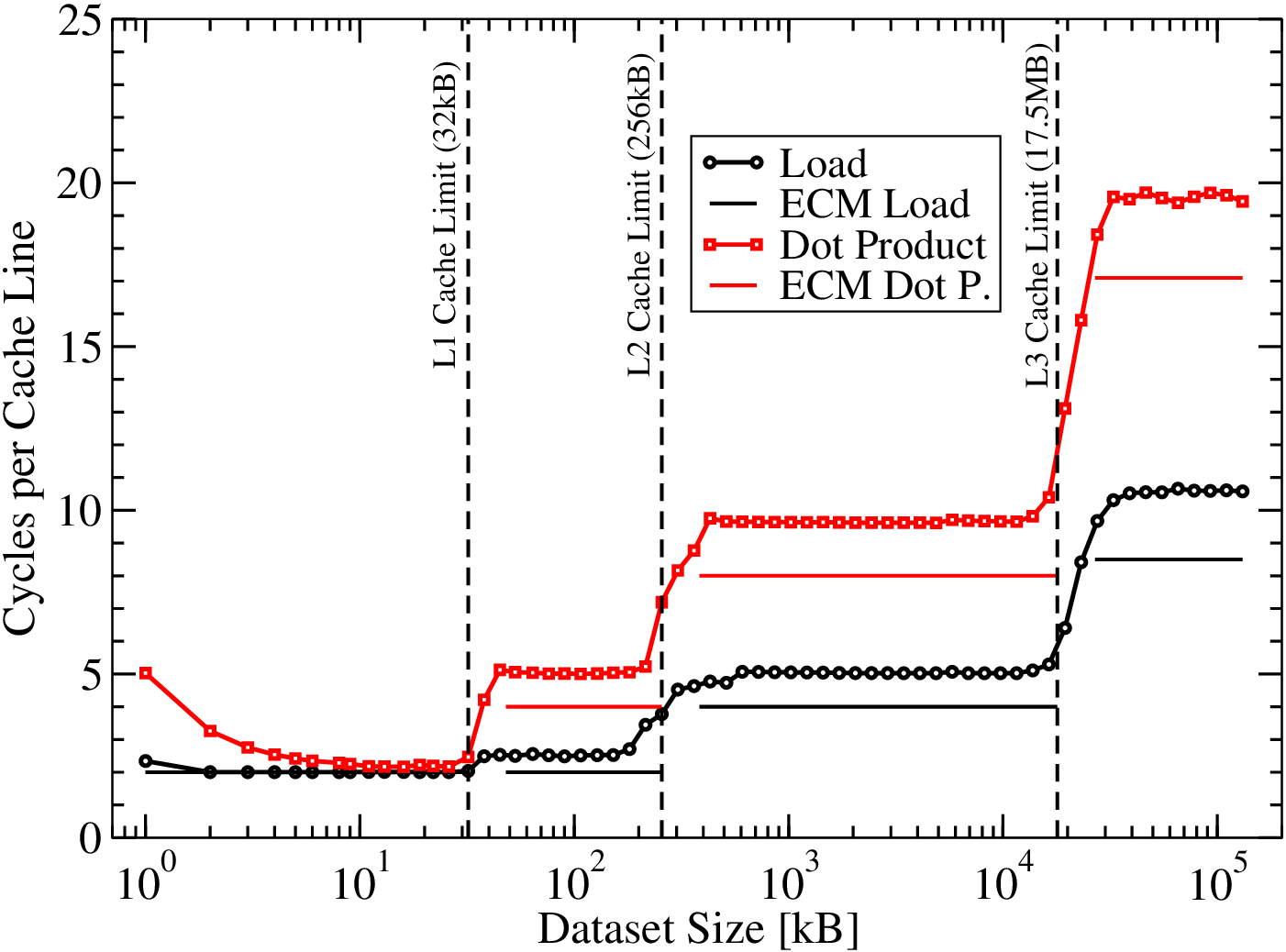}
    \caption{ECM predictions and measurement results for load and dot product kernels.}
    \label{fig:micro-loadddot}
\end{minipage}%
\hfill
\begin{minipage}{.49\textwidth}
    \centering
    \includegraphics[width=\linewidth]{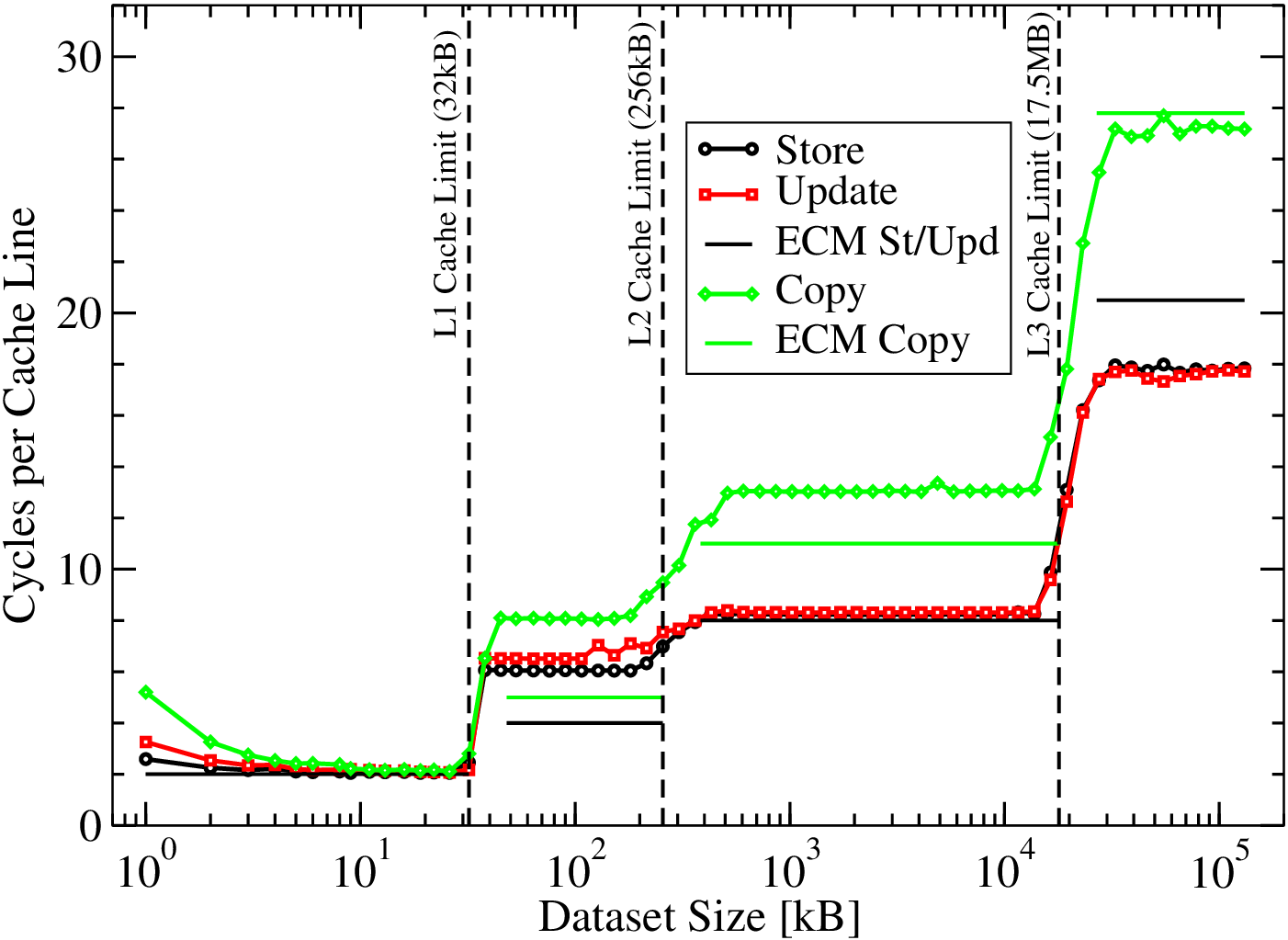}
    \caption{ECM predictions and measurement results for store, update, and copy kernels.}
    \label{fig:micro-stupco}
\end{minipage}
\end{figure}

Fig.~\ref{fig:micro-loadddot} illustrates ECM predictions and measurement
results for both the \textit{load} and \textit{ddot} benchmarks. While core
execution time for both benchmarks is two cycles as predicted by the model,
\textit{ddot} performance is slightly lower than predicted with data coming
from the L2 cache. The worse than expected L2 cache performance has been a
general problem with Haswell. In contrast to Haswell, Sandy and Ivy Bridge
delivered the advertised bandwidth of 32\,B/c \cite{sthw15}. On Haswell, in
none of the cases the measured L2 bandwidth could live up to the advertised
64\,B/c.  For the \textit{load} kernel, the performance in L2 is almost
identical to that with data residing in the L1 cache: this is because the CL
can theoretically be transfered from L2 to L1 a single cycle at 64\,B/c, which
is exactly the amount of slack that is the difference between
$T_\mathrm{OL}=2$\,c and $T_\mathrm{nOL}=1$\,c. In practise, however, we
observe a small penalty of 0.3\,c/CL, so again, we do can observe the specified
bandwidth of 64\,B/c.

As soon as the working set becomes too large for the core-local L2 cache, the
ECM prediction is slightly off. For kernels with a low number of cycles per CL
an empirically determined penalty for transferring data from off-core locations
was found to be one cycle per load stream and cache-level, e.g.  2\,c for the
\textit{ddot} benchmark with data residing in L3 and 4\,c with data from
memory. In all likelihood, this can be attributed to latencies introduced when
data is passing between different clock domains (e.g. core, cbox, mbox) that
cannot be entirely hidden for kernels with a very low core cycle count.

\subsection{Store, Update, and Copy}

Fig.~\ref{fig:micro-stupco} shows ECM predictions and measurements for the
\textit{store}, \textit{update}, and \textit{copy} kernels. With data in L1
cache, measurements for all three benchmarks match the prediction. In the L2
cache, measured performance is off about one cycle per stream: two cycles for
the \textit{store} and \textit{update} benchmarks, and four cycles for the
\textit{copy} benchmark. This means that it takes the data exactly twice as
long to be transfered than what would be the case assuming a bandwidth of
64\,B/c.

Measurements in L3 for the \textit{store} and \textit{update} kernel fit the
prediction. This suggests either that either overlap between transfers is
happening or some other undocumented optimization is taking place, as we would
normally expect the poor L2 performance to trickle down to L3 and memory
measurements (as is the case for the \textit{copy} kernel). Suspicion about
overlap or undocumented improvements is substantiated by better than expected
in-memory performance.

\subsection{Stream Triad and Sch\"onauer Triad}
\label{sec:res:triads}

\begin{figure}[tb]
\begin{minipage}{.485\textwidth}
    \centering
    \includegraphics[width=\linewidth]{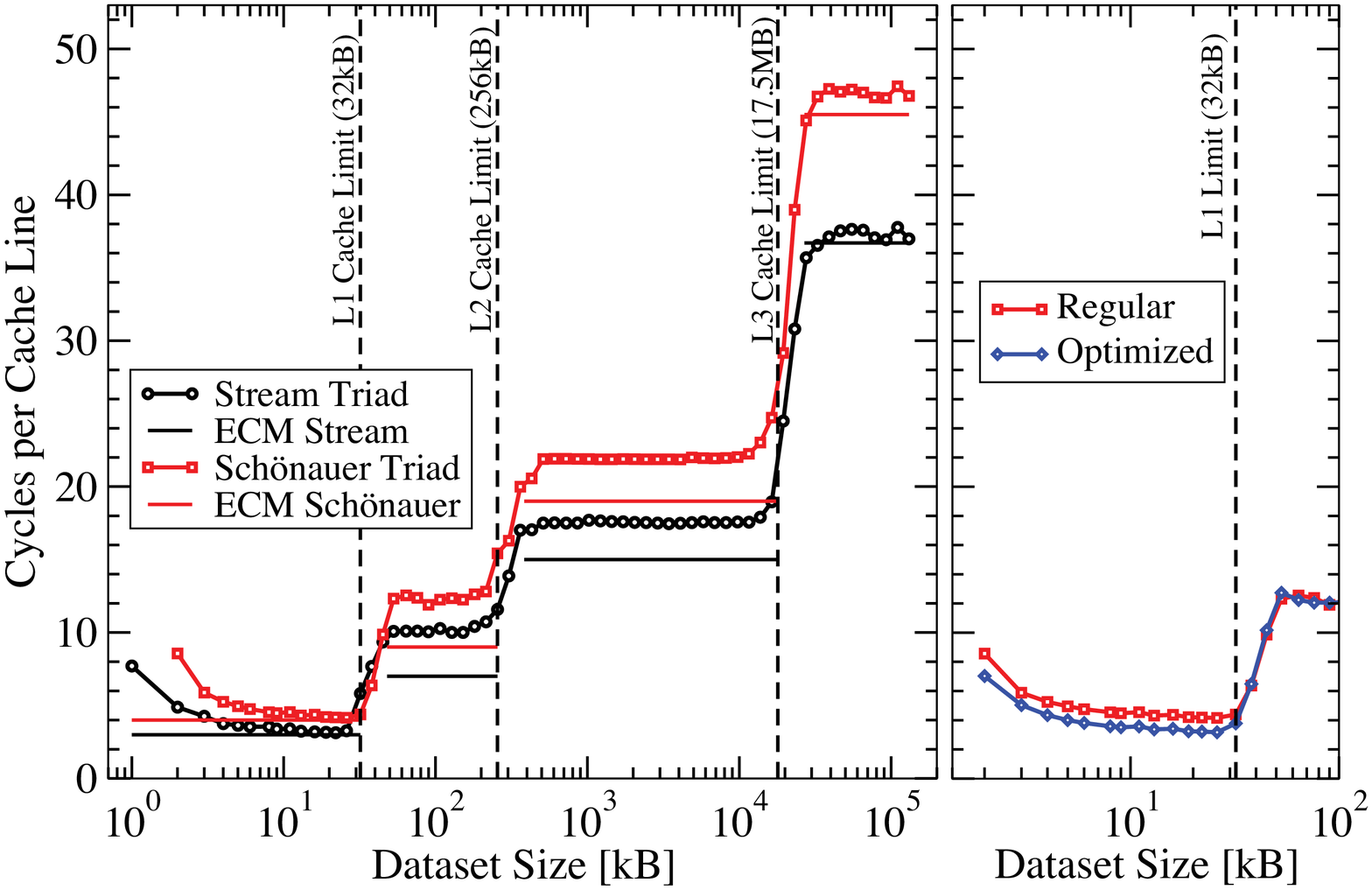}
    \caption{ECM predictions and measurement results for Stream and Sch\"onauer
    Triads (left) and comparison of naive and optimized Sch\"onauer Triad (right).}
    \label{fig:micro-triads}
\end{minipage}%
\hfill
\begin{minipage}{.50\textwidth}
    \centering
    \vspace{-0.34cm}
\lstset{
        breaklines=true,
        language=C,
        basicstyle=\footnotesize\ttfamily,
        numbers=left,
        numberstyle=\tiny,
        frame=tb,
        columns=fullflexible,
        showstringspaces=false
}
\begin{lstlisting}[caption={Shortened two-way unrolled, hand-optimized code for Sch\"onauer Triad. Eight-way unrolling used in real benchmark kernel.}, numbers=right, label=src:schoenauer_lea, captionpos=b, belowcaptionskip=4pt]{Name}
lea rbx, [r8+rax*8]
vmovapd ymm0, [rsi+rax*8]
vmovapd ymm1, [rsi+rax*8+32]
vmovapd ymm8, [rdx+rax*8]
vmovapd ymm9, [rdx+rax*8+32]
vfmadd231pd ymm0,ymm8,[rcx+rax*8]
vfmadd231pd ymm1,ymm9,[rcx+rax*8+32]
vmovapd [rbx], ymm0
vmovapd [rbx+32], ymm1
\end{lstlisting}
\end{minipage}
\end{figure}

Fig.~\ref{fig:micro-triads} shows model predictions and measurements for both
the Stream and Sch\"onauer Triads. The measurement fits the model's prediction
for data in the L1 cache. We observe the same penalty for data in the L2 cache.
This time, the penalty also propagates: measurement and prediction for data in
L3 is still off. The match of measurement and prediction for the in-memory case
suggests either overlap of transfers or other unknown optimization as was the
case before for the \textit{store}, \textit{update}, and \textit{copy}
kernels.

In addition, Fig.~\ref{fig:micro-triads} shows measurement results for the
naive Sch\"onauer Triad as it is currently generated by compilers (e.g. the
Intel C Compiler 15.0.1) and an optimized version that makes use of the newly
introduced simple \ac{AGU} on port~7. Typically, address calculations in
loop-unrolled streaming kernels require two steps: scaling and offset
computation. Both \ac{AGU}s on ports~2 and~3 support this addressing mode
called ``base plus index plus offset.'' The new simple \ac{AGU} can only
perform offset computations. However, it is possible to make use of this
\ac{AGU} by using one of the ``fast LEA'' units (which can perform
\textit{only} indexed and no offset addressing) to pre-compute an intermediary
address. This pre-computed address is fed to the simple \ac{AGU}, which can
then perform the still outstanding offset addition.  Using all three \ac{AGU}s,
it is possible to complete the eight addressing operations in three instead of
four cycles. The assembly code for this optimized version is shown in
Listing~\ref{src:schoenauer_lea}.
\subsection{Multi-Core Scaling and Cluster-on-Die Mode}

\begin{figure}[tb]
\begin{minipage}{.49\textwidth}
    \vspace{-\baselineskip}
    \centering
    \includegraphics[width=\linewidth]{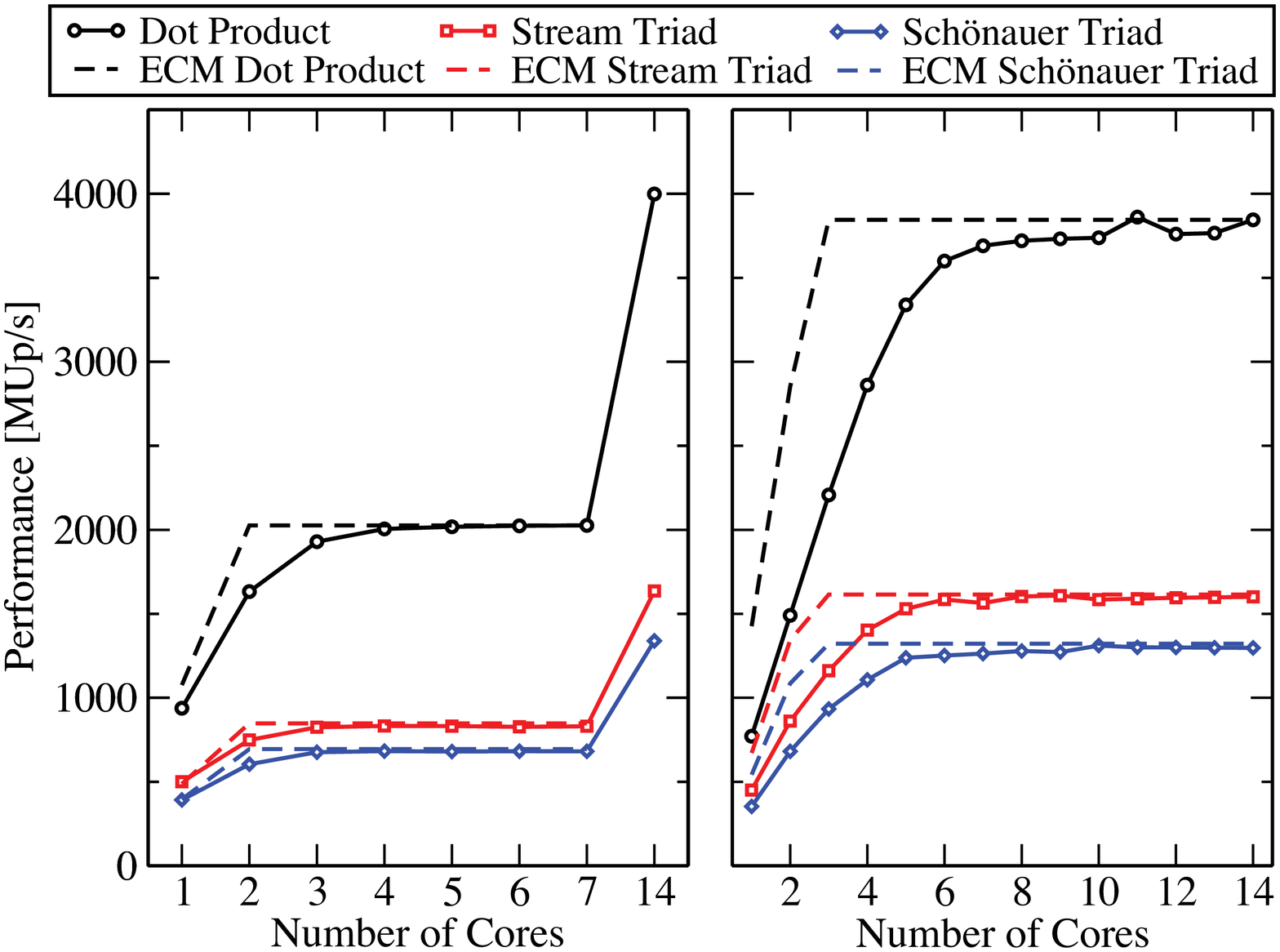}
    \caption{Core-Scaling using \ac{CoD} mode (left) and non-\ac{CoD} mode (right).}
    \label{fig:ecm-scaling}
\end{minipage}%
\hfill
\begin{minipage}{.49\textwidth}
    \centering
    \includegraphics[width=\linewidth]{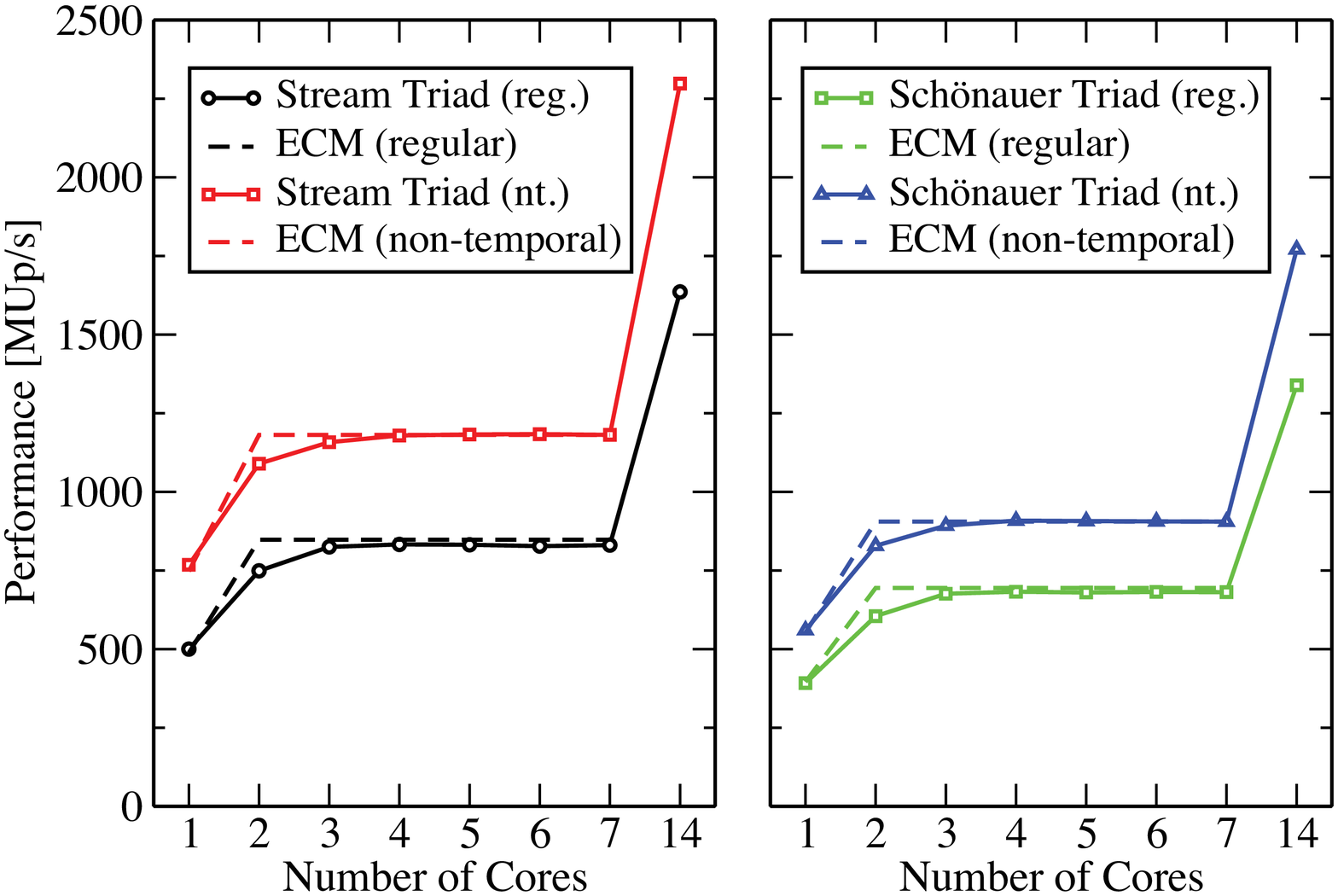}
    \caption{Performance using regular vs. non-temporal stores for Stream (left)
    and Sch\"onauer Triads (right).}
    \label{fig:ecm-scaling-nt}
\end{minipage}
\end{figure}

When using the ECM model to estimate multi-core performance, single-core
performance is scaled until a bottleneck is hit---which currently on Intel CPUs
is main memory bandwidth.  Fig.~\ref{fig:ecm-scaling} shows ECM predictions
along with actual measurements for the \textit{ddot}, \textit{Stream Triad},
and \textit{Sch\"onauer Triad} kernels using both \ac{CoD} and non-\ac{CoD}
modes. The L3-Memory CL transfer time
used for each prediction is based on the sustained bandwidth of the \ac{CoD}
respectively non-\ac{CoD} mode. While the measurement fits the prediction in
\ac{CoD} mode, we find a non-negligible discrepancy in non-\ac{CoD} mode. This
demonstrates how the ECM model can be used to uncover the source of performance deviations.
In non-\ac{CoD} mode, the kernel execution time is no
longer just made up of in-core execution and bandwidth-limited data transfers
as predicted by the model. Although we can only speculate, we attribute the
penalty cycles encountered in non-\ac{CoD} mode to higher latencies caused by
longer ways to the memory controllers: due to equal distribution of memory
requests, on average every second request has to go the ``long way'' across
rings, which is not the case in \ac{CoD} mode.

The measurements indicate that peak performance for both modes is nearly
identical, e.g. for \textit{ddot} performance saturates slightly below
4000\,MUp/s for non-\ac{CoD} mode while \ac{CoD} saturates slightly above the
4000 mark.  Although the plots indicate the bandwidth saturation point is
reached earlier in \ac{CoD} mode, this conclusion is deceiving. While it only
takes four cores to saturate the memory bandwidth of an memory domain, a single
domain is only using two memory controllers; thus, saturating chip bandwidth
requires $2\times4$ threads to saturate \textit{both} memory domains, the same
amount of cores it takes to achieve the sustained bandwidth in non-\ac{CoD}
mode.

\subsection{Non-Temporal Stores}

For streaming kernels with dataset sizes that do not fit into the \ac{LLC} it
is imperative to use non-temporal stores in order to achieve the best
performance.  Not only is the total amount of data to be transfered from memory
reduced by getting rid of \ac{RFO} stream(s), but in addition, data does not
have to travel through the whole cache hierarchy. On Haswell, non-temporal
stores are sent to the L1 cache by the core, just like regular stores; they do
however not update any entries in the L1 cache but are relayed to core-private
line fill buffers, from which data is transfered directly to main memory.

Fig.~\ref{fig:ecm-scaling-nt} shows the performance gain offered by
non-temporal stores. The left part shows the Stream Triad, which using regular
stores is made up of two explicit load streams for arrays \texttt{B} and {C}
plus a store and an implicit \ac{RFO} stream for array \texttt{A}. Looking at
transfered data volumes, we expect an performance increase by a factor of
$1.33\times$, because using non-temporal stores gets rid of the \ac{RFO}
stream, thereby reducing streams count from four to three.  However, the
measured speedup is higher: 1181 vs.  831\,MUp/s ($1.42\times$) using a single
memory domain respectively 2298 vs 1636\,MUp/s ($1.40\times$) when using a
full chip.  A possible explanation for this higher than anticipated speedup is
that we have observed the efficiency of the memory subsystem degrade with an
increasing number of streams. Vice verse, we could conclude that the efficiency
increases by getting rid of the \ac{RFO} stream.

A similar behavior is observed for the Sch\"onauer Triad. Data volume analysis
suggests a performance increase of $1.25\times$ (4 streams instead of 5).
However, the measured performance using non-temporal stores is 905 vs.
681\,GUp/s ($1.33\times$) using one memory domain resp. 1770 vs.  1339\,MUp/s
($1.32\times$) using a full chip.


\section{Conclusion}
This paper investigated new architectural features of the Intel Haswell
microarchitecture with regard to the execution of streaming loop kernels. It
demonstrated how to employ the \ac{ECM} model together with microbenchmarking
to quantify the efficiency of architectural features. On the example of a
comprehensive set of streaming loop kernels deviations from official
specifications as well as the overall efficiency was evaluated. Besides
incremental improvements and core related things Haswell addresses two main
areas: Energy efficiency and to provide low latency data access within the chip
while increasing the core count.  Sustained main memory bandwidth is no longer
impaired by the selection of low clock frequencies, enabling power savings of
more than 20\% respectively 10\% over the previous Sandy respectively Ivy
Bridge architectures. Uncore Frequency Scaling can further improve power
savings by more than 20\% for single-core workloads at no cost for data in
core-private caches respectively a small performance penalty with data
off-core.  The new Cluster-on-Die mode offers performance improvements for
single-threaded and parallel memory-bound codes, and has major benefits with
regard to latency penalties.

\bibliographystyle{splncs03}
\bibliography{pub}

\end{document}